\pdfoutput=1 
\documentclass[aps,twocolumn,prl,showpacs,amsmath,amssymb,10pt]{revtex4-1}
\usepackage{graphicx}
\usepackage{scrextend}
\usepackage{manfnt,pifont}
\usepackage{hyperref}
\usepackage{url}

\usepackage[normalem]{ulem}

\usepackage{url}
\usepackage[outline]{contour}
\usepackage{textgreek}

\usepackage{tikz}
\usetikzlibrary{shapes}

\usepackage{CJK}

\def\Dm{{\mathcal{D}}}
\def\Fm{{\mathcal{F}}}
\def\Hm{{\mathcal{H}}}
\def\Mm{{\mathcal{M}}}
\def\Pm{{\mathcal{P}}}

\newcommand\zb{{\bar z}}
\newcommand\Zb{{\bar Z}}
\newcommand\wh\widehat
\newcommand\Dh{{\widehat{\Delta}}}
\def\km{{k_{\mathrm m}}}

\begin{document}

\begin{CJK*}{GB}{}
\CJKfamily{gbsn}

\title{Defect two-point functions in 6d (2,0) theories}

\author{Junding Chen$^{a,b}$}
\author{Aleix Gimenez-Grau$^{c}$}
\author{Xinan Zhou$^{d,e}$}
\affiliation{$^{a}$CAS Key Laboratory of Theoretical Physics, Institute of Theoretical Physics, Chinese Academy of Sciences, Beijing 100190, China}
\affiliation{$^b$School of Physical Sciences, University of Chinese Academy of Sciences, No.19A Yuquan Road, Beijing 100049, China}
\affiliation{$^{c}$Institut des Hautes \'Etudes Scientifiques, 91440 Bures-sur-Yvette, France}
\affiliation{$^{d}$Kavli Institute for Theoretical Sciences, University of Chinese Academy of Sciences, Beijing 100190, China}
\affiliation{$^e$School of Physical Science and Technology, ShanghaiTech University,
Shanghai 201210, China.}

\begin{abstract}
\noindent We consider correlation functions in 6d $(2,0)$ theories of two $\frac{1}{2}$-BPS operators inserted away from a $\frac{1}{2}$-BPS surface defect. In the large central charge limit the leading connected contribution corresponds to sums of tree-level 
Witten diagram in AdS$_7\times$S$^4$ in the presence of an AdS$_3$ defect. We show that these correlators can be uniquely determined by imposing only superconformal symmetry and consistency conditions, eschewing the details of the complicated effective Lagrangian. We explicitly compute all such two-point functions. The result exhibits remarkable hidden simplicity.

\end{abstract}

\maketitle
\end{CJK*}

\section{Introduction}
\noindent Adding defects to QFTs greatly enriches the structure of theories. Such considerations have clear experimental motivations in representing impurities, domain walls and boundary effects in real-world systems. Formally, defects can be used to diagnose phases of theories \cite{Wilson:1974sk} and can also be interpreted as symmetry generators \cite{Gaiotto:2014kfa}. 

In the context of CFTs, introducing defects adds to the operator spectrum and OPE coefficients, commonly known as the CFT data, a new infinite set of numbers which defines the defect and its interaction with the bulk CFT. To access and extract these new data, it is most convenient to study correlation functions of local operators, but now in the presence of the non-local defect. This puts correlation functions at the center stage. For example, they are featured prominently in the bootstrap approach to defect CFTs \cite{Liendo:2012hy,Billo:2016cpy}, where nontrivial constraints on the defect CFT data are extracted from the crossing equation (see, e.g.,  \cite{Lemos:2017vnx,Liendo:2019jpu,Bissi:2018mcq,Kaviraj:2018tfd,Mazac:2018biw,Barrat:2022psm,Bianchi:2022ppi}). Meanwhile, it is also very important to be able to compute correlators in a given theory. So far most studies have focused on weak coupling, where standard techniques such as Feynman diagrams, $\epsilon$ expansion, and large $N$ expansion apply \cite{Die86a,cardy_1996,McAvity:1995zd,Gaiotto:2013nva,Cuomo:2021kfm,Cuomo:2022xgw,Giombi:2021uae,Giombi:2020rmc,Bianchi:2022sbz,Gimenez-Grau:2022ebb,Raviv-Moshe:2023yvq,Trepanier:2023tvb,Giombi:2023dqs}. In the strong coupling limit where AdS/CFT gives a useful dual description, although a number of results exist \cite{Chiodaroli:2016jod,Giombi:2017cqn,Drukker:2020swu,Ferrero:2021bsb,Barrat:2021yvp,Barrat:2022psm,Bianchi:2022ppi,Meneghelli:2022gps,Gimenez-Grau:2023fcy,Giombi:2023zte}, they are far from being systematic and comprehensive. By contrast, for CFTs without defects significant progress has been made in the modern analytic bootstrap program of holographic correlators, which was initiated in \cite{Rastelli:2016nze,Rastelli:2017udc} and has state of the art results at six points \cite{Alday:2023kfm} and at two loops \cite{Huang:2021xws,Huang:2023oxf} (see \cite{Bissi:2022mrs} for a recent review). It is natural to ask if the philosophy and techniques can be adapted to boost the study of correlators in defect CFTs. 

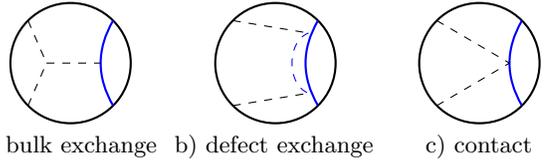
\begin{figure}
\begin{tabular}{c c c}
$\!\!\!$\begin{tikzpicture}[scale=0.8]
    \tikzstyle{every node}=[font=\scriptsize]
    \pgfmathsetmacro{\x}{sqrt(2)/2}
    \pgfmathsetmacro{\y}{sqrt(2)/2}
    \pgfmathsetmacro{\z}{-0.5}
    \pgfmathsetmacro{\r}{0.4}
    \draw [thick] (0,0) circle [radius=1];
    \draw [thick, blue] (\x,+\y) to[out=240,in=120] (\x,-\y);
    \draw [dashed] (-\x,+\y) -- (-\r,0);
    \draw [dashed] (-\x,-\y) -- (-\r,0);
    \draw [dashed] (-\r,  0) -- (-\z,0);
 \end{tikzpicture} &
$\!$\begin{tikzpicture}[scale=0.8]
    \tikzstyle{every node}=[font=\scriptsize]
    \pgfmathsetmacro{\x}{sqrt(2)/2}
    \pgfmathsetmacro{\y}{sqrt(2)/2}
    \pgfmathsetmacro{\rx}{0.55}
    \pgfmathsetmacro{\ry}{0.5}
    \draw [thick] (0,0) circle [radius=1];
    \draw [dashed]      (-\x,+\y) -- (\rx,\ry);
    \draw [dashed]      (-\x,-\y) -- (\rx,-\ry);
    \draw [dashed,blue] (\rx,\ry) to[out=200,in=160] (\rx,-\ry);
    \draw [thick, blue] (\x,+\y) to[out=240,in=120] (\x,-\y);
 \end{tikzpicture} &
$\;\;\;$\begin{tikzpicture}[scale=0.8]
    \tikzstyle{every node}=[font=\scriptsize]
    \pgfmathsetmacro{\x}{sqrt(2)/2}
    \pgfmathsetmacro{\y}{sqrt(2)/2}
    \pgfmathsetmacro{\z}{0.5}
    \draw [thick] (0,0) circle [radius=1];
    \draw [thick, blue] (+\x,+\y) to[out=240,in=120] (+\x,-\y);
    \draw [dashed]      (-\x,+\y) -- (\z,0);
    \draw [dashed]      (-\x,-\y) -- (\z,0);
  \end{tikzpicture} \\
  $\!\!\!\!\!$\text{a) bulk exchange} & 
  $\;$\text{b) defect exchange}$\;\;$ & 
 $\;\;\;$\text{c) contact}
 \end{tabular}
 \caption{The three types of tree-level Witten diagrams.}
 \label{fig:witten}
\end{figure}

This paper makes progress in this direction. We consider two-point correlation functions of $\frac{1}{2}$-BPS operators $S_k$ in the 6d $\mathcal{N}=(2,0)$ theory in the presence of a $\frac{1}{2}$-BPS surface defect $V$. The 6d $(2,0)$ theory is a strongly coupled SCFT with no Lagrangian description. The best way to attack this theory is via AdS/CFT where it is dual to M-theory on AdS$_7\times$S$^4$ and is approximated by eleven dimensional supergravity at large $N$. The operators $S_k$ are dual to an infinite Kaluza-Klein (KK) tower of scalar fields labelled by the KK level  $k=2,3,\ldots$ on S$^4$. Adding the defect amounts to introducing an M2-brane extended along an AdS$_3$$\subset$AdS$_7\times$S$^4$, which hosts localized degrees of freedom interacting with the bulk \footnote{One can also consider $M$ coincident M2-branes but this will only change the results of the paper by an overall factor.}. In such a setup, the two-point function of bulk operators is the simplest nontrivial observable. The standard large $N$ counting gives the following expansion
\begin{equation}\label{SSVexpand}
\begin{split}
    \langle S_{k_1} S_{k_2} V \rangle={}&\langle S_{k_1} S_{k_2}\rangle_{\rm free}+\frac{1}{N}\langle S_{k_1} V\rangle \langle S_{k_2} V\rangle \\
    {}&+\frac{1}{N^2}\langle S_{k_1} S_{k_2} V \rangle_{\rm tree}+\mathcal{O}(N^{-3})\; ,
    \end{split}
\end{equation}
where we have explicitly extracted the $N$ dependence. The first two terms are disconnected contributions and correspond to the free AdS propagator and the product of one-point functions respectively. The leading nontrivial contribution is the connected part $\langle S_{k_1} S_{k_2} V \rangle_{\rm tree}$ which corresponds to a sum of tree-level Witten diagrams in the presence of a defect, see Figure \ref{fig:witten}. These tree-level correlators are the main focus of this paper. The traditional diagrammatic approach, albeit viable in principle, requires inputting the  precise details of the AdS effective Lagrangian which are difficult to obtain. The main result of our paper is to show that symmetry principles render this unnecessary and allow us to eschew all such details. We will uniquely determine all defect two-point functions with arbitrary $k_1$, $k_2$ from superconformal symmetry and consistency conditions. Moreover, we  uncover remarkable simplicity by presenting a compact expression for general correlators.

\section{Kinematics}
\noindent
The operators $S_k$ have protected conformal dimensions $\Delta_{S_k} = 2k$ and transform in rank-$k$ symmetric traceless representations of the R-symmetry group $SO(5)$. To keep track of R-symmetry, we contract the indices with null polarization vectors $u^I$
\begin{align}
 S_k(x,u) = S_{I_1 \ldots I_k} u^{I_1} \ldots u^{I_k} \, , \qquad
 u \cdot u = 0 \; .
\end{align}
The defect operator $V$ breaks $OSp(8^*|4)$ superconformal symmetry of the original theory into $[OSp(4^*|2)]^2$. In particular, $V$ divides the coordinates into the parallel part $x^{a=1,2}$ and the transverse part $x^{i=3,\ldots,6}$, which are respectively acted on by the defect conformal group $SO(2,2)$ and the orthogonal $SO(4)$ rotations. It also breaks the R-symmetry as $SO(5)\to SO(4)$. The  embedding of $SO(4)\subset SO(5)$ can be captured by a polarization vector $\theta$ with $\theta^2 = 1$. Then the defect two-point function can be written as \footnote{The normalization is ${\langle S_k S_k\rangle}_{\rm free}=(u_1\cdot u_2)^k/x_{12}^{4k}$  in (\ref{SSVexpand}).}
\begin{align}\label{d2ptfun}
 \langle S_{k_1} S_{k_2} V \rangle
 = \frac{(u_1 \cdot \theta)^{k_1} (u_2 \cdot \theta)^{k_2}}
        {|x_1^i|^{2k_1} |x_2^i|^{2k_2}}
   \Fm(z,\zb,\sigma) \; ,
\end{align}
where the $z,\zb$ cross ratios are defined by
\begin{equation}
    \frac{x_{12}^2}{|x_1^i||x_2^i|}=\frac{(1-z)(1-\bar{z})}{\sqrt{z\bar{z}}}\;,\quad \frac{x_1^jx_2^j}{|x_1^i||x_2^i|}=\frac{z+\bar{z}}{2\sqrt{z\bar{z}}}\;,
\end{equation}
and the R-symmetry cross ratio is
\begin{align}
\label{Rcrossratiosigma}
 \sigma
 = \frac{u_1 \cdot u_2}{(u_1 \cdot \theta)(u_2 \cdot \theta)}
 = - \frac{(1-\omega)^2}{2\omega} \; .
\end{align}
From the definition, it is clear that the correlator is a polynomial in $\sigma$
\begin{align}
 \Fm(z,\zb,\sigma) = \sum_{n=0}^{\km} \sigma^n \Fm_n(z,\zb) \; ,
\end{align}
where $\km\equiv\min{(k_1,k_2)}$.
In (\ref{d2ptfun}) we have only exploited the bosonic part of the unbroken superconformal symmetry. The fermionic generators impose further constraints known as the superconformal Ward identity \cite{Meneghelli:2022gps}
\begin{align}
\label{eq:wi}
 (\partial_z + \partial_\omega) \Fm(z,\zb,\omega)|_{z=\omega}
 = 0 \; ,
\end{align}
together with its $z \leftrightarrow \zb$ counterpart. Equivalently, we have 
\begin{align}
 \Fm(z,\zb,\zb) = \zeta(z) \; , \quad
 \Fm(z,\zb,z) = \zeta(\zb) \; ,
\end{align}
which is a consequence of the chiral algebra and $\zeta(z)$ is the chiral correlator \cite{bllprv13,brv14,Meneghelli:2022gps}. The most general solution reads
\begin{equation}\label{solscfWardid}
  \Fm(z,\zb,\omega)=\Fm_{\rm prot}(z,\zb,\omega)+\mathrm{R}\,  \Hm(z,\zb,\omega)\;,
\end{equation}
where
\begin{equation}\label{Rfactor}
    \mathrm{R}=\frac{(z-\omega)(\zb-\omega)(z-\omega^{-1})(\zb-\omega^{-1})}{z \zb }\; ,
\end{equation}
and $\Fm_{\rm prot}$ is the protected part
\begin{equation}
   \Fm_{\rm prot}(z,\zb,\omega)=  \frac{(z-\omega)(z-\omega^{-1})}{(z-\zb)(z-\zb^{-1})} \zeta(z)
 + (z \leftrightarrow \zb)\; .
\end{equation}
The function $\Hm$ is called the {\it reduced correlator} and it is a polynomial in $\sigma$ of degree $\km-2$, given explicitly by
\begin{align}
\nonumber \Hm
= \sum_{n=2}^\km 
  \frac{\sigma^n (Z-\Zb)+Z^n (\Zb-\sigma)+\Zb^n (\sigma-Z)}
       {4(\sigma-Z) (\sigma-\Zb) (Z-\Zb)} \Fm_n(z,\zb) \; ,
\end{align}
where $Z = -(1-z)^2/(2z)$, and similarly for $\Zb$ with $z$ replaced by $\bar{z}$.

\section{Bootstrap algorithm}
\noindent Following AdS/CFT, the leading large $N$ contribution to the connected defect two-point functions can be computed as a finite sum of tree-level Witten diagrams. In principle, this can be done by expanding the defect effective action in AdS to the quadratic order and extracting the Feynman rules, as has been attempted for the Wilson line case \cite{Gimenez-Grau:2023fcy}. However, in practice this is very cumbersome and difficult to follow through due to subtleties noted in \cite{Gimenez-Grau:2023fcy} for contact interactions. A more efficient strategy is to keep the coefficients  unfixed and fix them using superconformal symmetry. Such a strategy, originally formulated in \cite{Rastelli:2016nze,Rastelli:2017udc} in the context of correlators in CFTs without defects and termed the position space approach, was applied to compute Wilson line two-point functions in \cite{Gimenez-Grau:2023fcy}. Here we extend it to surface defects. 

Our starting point is the following ansatz
\begin{equation}
    \mathcal{F}_{\rm ansatz}=    \mathcal{F}_{\rm exchange}^{\rm bulk}+\mathcal{F}_{\rm exchange}^{\rm defect}+\mathcal{F}_{\rm contact}\;,
\end{equation}
with the corresponding diagrams shown in Figure ~\ref{fig:witten}. The exchange contribution is divided into a bulk part
\begin{equation}
  \mathcal{F}_{\rm exchange}^{\rm bulk}=\sum_{\mathcal{X}}\mu_{\mathcal{X}}E_{2k_1,2k_2}^{\Delta_{\mathcal{X}},\ell_{\mathcal{X}}}(z,\bar{z}) h_{k_1k_2}^{R_{\mathcal{X}}}(\sigma)\;, 
\end{equation}
and a defect part
\begin{equation}
  \mathcal{F}_{\rm exchange}^{\rm defect}=\sum_{\mathcal{Y}}\widehat{\mu}_{\mathcal{Y}}\widehat{E}_{2k_1,2k_2}^{\widehat{\Delta}_{\mathcal{Y}},s_{\mathcal{Y}}}(z,\bar{z}) \widehat{h}_{R_{\mathcal{Y}}}(\sigma)\; , 
\end{equation}
where we sum over all possible exchanged fields $\mathcal{X}$ and $\mathcal{Y}$ with unknown coefficients $\mu_{\mathcal{X}}$ and $\widehat{\mu}_{\mathcal{Y}}$. The bulk exchange Witten diagram $E_{2k_1,2k_2}^{\Delta_{\mathcal{X}},\ell_{\mathcal{X}}}$ has internal conformal dimension $\Delta_{\mathcal{X}}$ and Lorentz spin $\ell_{\mathcal{X}}$ and the defect channel exchange Witten diagram $\widehat{E}_{2k_1,2k_2}^{\widehat{\Delta}_{\mathcal{Y}},s_{\mathcal{Y}}}$ has dimension $\widehat{\Delta}_{\mathcal{Y}}$ and transverse spin $s_{\mathcal{Y}}$. The R-symmetry polynomials $h_{k_1k_2}^{R_{\mathcal{X}}}$ and $\widehat{h}_{R_{\mathcal{Y}}}$ capture the exchange of irreducible representations in the bulk and defect channels with R-symmetry charges $R_{\mathcal{X}}$ and $R_{\mathcal{Y}}$ respectively \footnote{More precisely, the bulk field transforms in the rank-$R_{\mathcal{X}}$ symmetric traceless representation of the original $SO(5)$ R-symmetry group while the defect field transforms in the rank-$R_{\mathcal{Y}}$ representation with respect to the unbroken $SO(4)$ R-symmetry.}. They can be obtained by solving quadratic Casimir equations  
\begin{align}
 h^{k}_{k_1 k_2}(\sigma)
&= \sigma^{\frac{k_1+k_2-k}{2}} {}_2F_1\!\left(\frac{k_{12}-k}{2},\frac{k_{21}-k}{2};-k-\frac{1}{2};\frac{\sigma }{2}\right) , \notag \\
 \wh h_{k}(\sigma) 
&= \sigma^k \, {}_2F_1\!\left(-k-\frac{1}{2},-k;-2 k-1;\frac{2}{\sigma} \right) ,
\end{align}
where $k_{ij}=k_i-k_j$. The set of fields which can be exchanged is finite and is constrained by a number of conditions.
First of all, R-symmetry selection rules impose constraints on what representations can appear in the exchanges. Looking at the spectrum of the theory, this already ensures the finiteness of the set. Second, all exchanges must be {\it non-extremal}. It means $\Delta_{\mathcal{X}}-\ell_{\mathcal{X}}<2k_1+2k_2$ in the bulk channel and $\widehat{\Delta}_{\mathcal{Y}}-s_{\mathcal{Y}}<2\min\{k_1,k_2\}$ in the defect channel. This condition arises from the vanishing of extremal couplings which is needed to have a finite effective Lagrangian. Third, $\ell_{\mathcal{X}}$ is restricted to even spins (i.e., $\ell_{\mathcal{X}}=0,2$) because spinning bulk fields are coupled to the components of the metric transverse to the defect (e.g. $\int_{AdS_3} g^{ij}\phi_{ij}$). Finally, for any $k_1$, $k_2$, $\mathcal{Y}$ in fact can only be two fields with $\widehat{\Delta}_{\mathcal{Y}}=2$, $s_{\mathcal{Y}}=0$ and $\widehat{\Delta}_{\mathcal{Y}}=3$, $s_{\mathcal{Y}}=1$. This is due to the fact that the defect is an AdS$_3$ inside AdS$_7\times$S$^4$ and there is no internal manifold to generate infinite KK modes. The  spectrum of exchanged fields is summarized in Table \ref{tablespectrum}.
\begin{table}[h]
\begin{center}
{\renewcommand{\arraystretch}{1.3}
\begin{tabular}{c c c c || c c c c }
~Bulk~ & $\Delta$ & $~~\ell~~$  & $\,$R charge$\,$ & ~Defect & ~~$\widehat{\Delta}$~~ &  $~~s~~$  & $\,$R charge$\,$ \\
 \hline
 \hline
$S_k$ & $2k$ &  0  & $k$ & $\phi$ &2 &0 & 1\\
$\Phi^{\mu\nu}_k$ & $2k+2$ &  2  & $k-2$ & $\rho^i$ &3 &1 & 0 \\ 
 $T_k$ & $2k+4$ & 0 & $k-4$ & & & & \\ 
\end{tabular}}
 \caption{Spectrum of exchanged fields in bulk and defect channels.}
\label{tablespectrum}
\end{center}
\end{table}

The ansatz also contains a contact part which we parameterize as 
\begin{equation}
   \mathcal{F}_{\rm contact}=\sum_{n=0}^{\km} \bar\mu_n \sigma^n C_{2k_1,2k_2}(z,\bar{z})\;.  
\end{equation}
Here $C_{\Delta_1 \Delta_2}$ is the zero-derivative contact Witten diagram.
Note that we have included in the ansatz all possible R-symmetry structures. However, we do not include contact Witten diagrams with more derivatives in the vertex. Such contact Witten diagrams are more dominant than the exchange Witten diagrams in the Regge limit, which is not expected.

The next step is to evaluate the Witten diagrams in the ansatz. For our theory, the spectrum is such that both the bulk and defect exchange Witten diagrams can be expressed as a finite sum of contact Witten diagrams, and the contact diagrams are also known explicitly \cite{Rastelli:2017ecj,Gimenez-Grau:2023fcy}. For convenience, we collect all these results in the  Supplemental Material. Using the explicit expressions, we can evaluate the full ansatz for any $k_1$, $k_2$. It is then straightforward to impose the superconformal Ward identity (\ref{eq:wi}) and we find that all unknown parameters in the ansatz can be solved up to an overall constant. Since the parameters can be interpreted as OPE coefficients and the same  coefficients appear in different correlators, by considering two-point functions with different $k_1$, $k_2$ we can further reduce the overall constants to just one for $k_1=k_2=2$. This can be fixed in terms of central charges as in \cite{Meneghelli:2022gps}. The bootstrap algorithm then completely determines all the defect two-point functions.

\section{Defect two-point functions}
\noindent The result of the bootstrap calculation becomes more illuminating when expressed in terms of the Polyakov-Regge superblocks defined in \cite{Gimenez-Grau:2023fcy}
\begin{align}
 \Pm_k
&= P^{2 k,0}_{2k_1,2k_2} h^k_{k_1k_2}
 + \alpha_k P^{2 k+2,2}_{2k_1,2k_2} h^{k-2}_{k_1 k_2} 
 + \beta_k  P^{2 k+4,0}_{2k_1,2k_2} h^{k-4}_{k_1 k_2} \, , \notag \\
 \wh \Pm
&= \wh P_{2k_1,2k_2}^{2,0} \wh h_1
 + \frac{1}{2} \wh P_{2k_1,2k_2}^{3,1} \wh h_{0} \; .
 \label{eq:superblock}
\end{align}
Here $P$, $\widehat{P}$ are rescaled $E$, $\widehat{E}$ and with contact Witten diagrams added in the bulk channel case to improve the Regge behavior \cite{Mazac:2018biw,Gimenez-Grau:2023fcy} (see Supplemental Material for details). The coefficients $\alpha_k$ and $\beta_k$ are fixed by the Ward identity \eqref{eq:wi} to be 
\begin{align}
\nonumber \alpha_k = \frac{(k^2-k^2_{12}) ((k+1)^2 - k^2_{12})}{8 (2 k-1) (2 k+1)^2 (2 k+3)} \; , \quad
 \beta_k = \alpha_{k} \alpha_{k-2} \; .
\end{align}
Each Polyakov-Regge superblock corresponds to the contribution from exchanging a particular supermultiplet. In terms of these building blocks, the tree-level defect two-point functions read
\begin{equation}\label{eq:pos-res}
    \Fm= \sum_{k} \lambda_{k_1k_2k} a_{k} \Pm_{k}
  + b_{k_1\Dm} b_{k_2\Dm} \wh\Pm
  + c_{k_1} c_{k_2} (1-2\sigma) C_{2k_1,2k_2}\,,
\end{equation}
where the sum over $k$ runs from $k_{\text{min}} =|k_{12}|+2$ to $k_{\text{max}} =k_1+k_2-2$ in steps of two.
Note that the contact part becomes particularly simple. To manifest the physical meaning of the coefficients, let us factor out in the bulk channel the bulk three-point OPE coefficients \cite{Corrado:1999pi,Bastianelli:1999en}
\begin{align}
 \lambda_{k_1k_2k_3}
&=\frac{2^{\Sigma-2}
        \Gamma \! \left(\frac{\Sigma}{2}\right)}
       {\pi^{3/2}} 
  \prod_{i=1}^3 \frac{\Gamma \! \left(\frac{\Sigma-2k_i+1}{2}\right)}{\sqrt{\Gamma (2 k_i-1)}} \; ,
\end{align}
with $\Sigma = k_1+k_2+k_3$. Our bootstrap calculation gives the defect OPE coefficients
\begin{align}
\nonumber a_k
 = \frac{1}{k} b_{k\Dm}
 = \frac{(k-1)(2k-1)}{2^{k-\frac{1}{2}} \sqrt{\pi} \, c_k}
 = \frac{\Gamma (k)}{\sqrt{2^{k} \Gamma (2 k-1)}} \; .
\end{align}
The $a_k$ were computed in \cite{Corrado:1999pi}\footnote{Our result for $a_k$ agrees with \cite{Corrado:1999pi} up to a simple factor. A similar mismatch was also observed in \cite{Bastianelli:1999en} for three-point functions.} while the $b_{k\Dm}$ are new predictions. Taking $k_1=k_2=2$ our result (\ref{eq:pos-res}) reproduces the special case computed in \cite{Meneghelli:2022gps}.

From the holographic two-point function we can also extract the chiral algebra correlator. This is achieved by setting $\omega=\bar{z}$. We find that the defect two-point function reduces to the following simple meromorphic function
\begin{equation}\label{chiralzeta}
 \zeta(z) 
 =  \frac{1}{2}\, b_{k_1\mathcal{D}} b_{k_2\mathcal{D}}
   \sum_{i=1}^{\km-1} C_i \, (2 Z)^{-i} \; ,
\end{equation}
where $C_i=\frac{1}{i+1}\binom{2i}{i}$ is the Catalan number. The chiral algebra of the 6d $(2,0)$ theory is conjectured to be the $\mathcal{W}_N$ algebra, where $S_k$ inserted on the chiral algebra plane is mapped to the $k$-th generator $W_k$ \cite{Meneghelli:2022gps,brv14}. Meanwhile, the defect is mapped to two vertex operators $\mathrm{V}(0)$, $\bar{\mathrm{V}}(\infty)$ inserted at zero and infinity. Then $\zeta(z)$ is the four-point function $\langle\bar{\mathrm{V}}(\infty) W_{k_1}(z) W_{k_2}(1) \mathrm{V}(0)\rangle$. The correlator $\zeta(z)$ can also be computed purely in the 2D chiral CFT from the knowledge of the OPE. One starts with a rational function ansatz of which the singularities and their strengths are dictated by the OPE. This ansatz can be written in the form of (\ref{chiralzeta}) thanks to the $z\leftrightarrow 1/z$ invariance with unfixed coefficients. These coefficients can then be solved by comparing the singularities with the prediction from the OPE (see, e.g.,  \cite{Rastelli:2017ymc,Behan:2021pzk,Meneghelli:2022gps}). We will not pursue this calculation further here. However, we point out that the ansatz can also be more efficiently fixed by requiring the small $z$ expansion contains no other powers between the leading term $z$ and the next term $z^\km$. This is due to the fact that only one supermultiplet of fields appear in the defect exchange Witten diagrams and the spectrum of protected operators has a gap.

\section{Mellin space}
\noindent
A natural language for holographic correlators is Mellin space \cite{Mack:2009mi,Penedones:2010ue}, where their  analytic structure is drastically simplified and the scattering amplitude nature becomes manifest. The Mellin formalism can also be extended to correlators in CFTs with boundaries and defects \cite{Rastelli:2017ecj,Goncalves:2018fwx}, where  Mellin amplitudes are interpreted as form factors of particles scattering with an extended object. The Mellin representation for defect two-point functions is given by \cite{Goncalves:2018fwx}
\begin{align}
\label{Mellindef}
&\mathcal{F}=\int \frac{d \delta \, d\gamma}{(2\pi i)^2}  
\xi^{-\delta} \chi^{-\gamma+\delta}  \mathcal{M}(\delta,\gamma)\Gamma_{k_1k_2}(\delta,\gamma)\;,
\end{align}
where we recombine the cross ratios $z$, $\bar{z}$ into
\begin{equation}
    \xi=\frac{(1-z)(1-\bar{z})}{\sqrt{z\bar{z}}}\;,\quad \chi=\frac{z+\bar{z}}{\sqrt{z\bar{z}}}\;.
\end{equation}
The dynamical information is encoded in the Mellin amplitude $\mathcal{M}$ and
\begin{align}
   \Gamma_{k_1k_2}= \Gamma(\delta)\Gamma(\gamma-\delta)\prod_{i=1}^2\Gamma\!\left(\frac{2k_i-\gamma}{2}\right)\,,
\end{align}
is a factor included as part of the definition. In this representation,  contact Witten diagrams have constant Mellin amplitudes and  the Mellin amplitudes of exchange diagrams have only simple poles. Their explicit expressions can be found in \cite{Gimenez-Grau:2023fcy} (also in Supplemental Material) and we can obtain the Mellin amplitudes of two-point functions by translating \eqref{eq:pos-res} diagram by diagram. 
However, it turns out that the most compact way to express the result is to use the reduced Mellin amplitude, defined by
\begin{align}
\label{strippedreducedmellindef}
&\mathcal{H}=\int \frac{d \delta \, d\gamma}{(2\pi i)^2}  
\xi^{-\delta} \chi^{-\gamma+\delta}  \widetilde{\mathcal{M}}(\delta,\gamma)\widetilde{\Gamma}_{k_1k_2}(\delta,\gamma)\;,
\end{align}
where we extract a different Gamma factor
\begin{align}
   \widetilde{\Gamma}_{k_1k_2}= \Gamma(\delta+1)\Gamma(\gamma-\delta)\prod_{i=1}^2\Gamma\!\left(\frac{2k_i+2-\gamma}{2}\right)\,.
\end{align}
Similar to the case of four-point functions in AdS$_5\times$S$^5$ \cite{Rastelli:2016nze,Rastelli:2017udc}, the protected part in (\ref{solscfWardid}) does not contribute to the Mellin amplitude. Comparing the definitions (\ref{Mellindef}) and (\ref{strippedreducedmellindef}),  we find that factor $\mathrm{R}$ in (\ref{Rfactor})
\begin{equation}
    \mathrm{R}=\xi^2 +2  \xi \chi \sigma+2 \chi ^2 \sigma -8\sigma+4 \sigma ^2\;,
\end{equation}
acts as a difference operator $\widehat{\mathrm{R}}$ in Mellin space which can be obtained by promoting each monomial as
\begin{equation}
\begin{split}
    \nonumber \widehat{\xi^m\chi^n}\circ\widetilde{\mathcal{M}}(\delta,\gamma)={}& \widetilde{\mathcal{M}}(\delta+m,\gamma+m+n)\\
    \times {}&\frac{\widetilde{\Gamma}_{k_1k_2}(\delta+m,\gamma+m+n)}{\Gamma_{k_1k_2}(\delta,\gamma)}\;.    
\end{split}
\end{equation}
Then the full Mellin amplitude is related to the reduced Mellin amplitude by
\begin{equation}
    \mathcal{M}(\delta,\gamma)=\widehat{\mathrm{R}}\circ\widetilde{\mathcal{M}}\;.
\end{equation}
From the position space result (\ref{eq:pos-res}), we find that the reduced correlator can always be written as a finite sum of contact Witten diagrams. It is then straightforward to translate the reduced correlators into Mellin space and we find that the reduced Mellin amplitudes admit a remarkably simple form with only simultaneous poles
\begin{align}
\widetilde{\mathcal{M}}(\delta,\gamma,\sigma)= \!\sum_{i=1}^{2\km-2} \sum_{j=2}^{\km} \frac{\mathfrak{R}_{ij}(\sigma)}{(\delta-i)(\gamma-2j)}\;,
\end{align}
where the residues are given by
\begin{align}
\nonumber \mathfrak{R}_{ij}(\sigma)=\!\sum_{m=\lfloor \frac{i}{2}\rfloor }^{\min{(i,j-1)}} \frac{b_{k_1\Dm} b_{k_2\Dm} (-1)^{i} \binom{m}{i-m}(2\sigma)^{m-1}}{2j! m! (k_1-j)!(k_2-j)! (j-m-1)!}\;.
\end{align}

\section{Discussion}
\noindent
In this paper, we performed a systematic bootstrap analysis of two-point functions of $\frac{1}{2}$-BPS operators in the 6d $(2,0)$ theory in the presence of a surface defect and obtained all tree-level correlators with arbitrary KK levels. There are many interesting future research directions. First, our result for two-point functions is surprisingly simple, especially when written in Mellin space. This is highly reminiscent of tree-level four-point functions of IIB supergravity in AdS$_5\times$S$^5$, where a similar unexpected simplicity led to the discovery of higher dimensional conformal symmetries in a number of models \cite{Caron-Huot:2018kta,Rastelli:2019gtj,Alday:2021odx,Abl:2021mxo}. While the same symmetry is clearly ruled out by the explicit four-point correlators \cite{Alday:2020lbp}, we nevertheless expect that some form of higher dimension structure should exist in all models to organize correlators of different KK modes. Unfortunately, no such organizing principles are known at the moment. The simple form of the defect two-point functions provides an ideal starting point for exploring new structures. Second, the two-point functions encode a wealth of defect CFT data. It would be interesting to extract this data and use it to compute loop corrections by extending the AdS unitarity method \cite{Aharony:2016dwx} to the defect case. These loop-level correlators will allow us to probe M-theory corrections beyond supergravity. Third, while in this paper we have restricted our attention to a specific model, the same strategy can be applied to compute defect two-point functions in an array of other setups. Prime targets include  surface defects in 4d $\mathcal{N}=4$ SYM and line defects in 3d ABJM theories, to name just a few. From examining these models, it would be very interesting if we can understand the general structure of these holographic defect correlators by writing down an interpolating formula parameterized by the spacetime dimension and defect dimension, similar to the four-point function case with no defects \cite{Alday:2020dtb}. Finally, we can also extend our bootstrap program to encompass higher-point defect correlators. Applying our techniques, a reasonable goal is to compute three-point functions with one bulk and two defect operators or two bulk and one defect operators which have the simplest kinematics. A study of the relevant Witten diagrams has been initiated in \cite{Chen:2023oax}.

\vspace{2.5cm}

The work of J.C. and X.Z. is supported by funds from University of Chinese Academy of Sciences (UCAS), funds from the Kavli Institute for Theoretical Sciences (KITS), the Fundamental Research Funds for the Central Universities, and the NSFC Grant No. 12275273.
AGG is supported by the Simons Foundation by grants 915279 (IHES) and 733758 (Bootstrap Collaboration).

\appendix

\section{Supplemental Material}

\noindent The contact Witten diagram in position space is given by
\begin{align}\label{Ck1k2}
& C_{\Delta_1 \Delta_2}
  = \frac{\pi^{3/2}}
         {2^{\Delta_1+\Delta_2}}
    \frac{\Gamma\!\left(\frac{\Delta_1+\Delta_2-2}{2}\right)}
         {\Gamma\!\left(\frac{\Delta_1+\Delta_2+1}{2}\right)} \notag \\
& \quad 
  \times {}_2F_1\!\left(\Delta_1,\Delta_2;\frac{\Delta_1+\Delta_2+1}{2};-\frac{\xi+\chi-2}{4} \right) \,.
\end{align}
The exchange Witten diagrams can be expressed as finite sums of contact Witten diagrams
\begin{align}
 E_{\Delta_1\Delta_2}^{\Delta,0}
&= \sum _{n=1}^{\bar\Delta}
   \frac{\left(1-\bar\Delta\right)_{n-1} 
    \left(4-\Delta-\bar\Delta\right)_{n-1}
  }{4 \xi ^n (1-\Delta_1)_n (1-\Delta_2)_n} C_{\Delta_1-n,\Delta_2-n}  \; , \notag \\
 \wh E_{\Delta_1\Delta_2}^{\Dh,0}
&= \!\sum_{n=1}^{\frac{\Delta_1-\Dh}{2}} 
   \frac{\left(\frac{\Dh+2-\Delta_1}{2}\right)_{n-1} 
         \left(\frac{4-\Dh-\Delta_1}{2}\right)_{n-1}}
        {4 (1-\Delta_1)_{2 n}}
    C_{\Delta_1-2 n,\Delta_2} \; , \notag \\
 \wh E^{\wh\Delta,1}_{\Delta_1\Delta_2}
&= 2 \Delta_1 \Delta_2 \chi \,
   \wh E^{\wh\Delta,0}_{\Delta_1+1,\Delta_2+1} \; ,
 \label{eq:spin-one-shift}
\end{align}
where we introduced $\bar \Delta = \frac{\Delta_1+\Delta_2-\Delta}{2}$.
The spin-two diagram $E_{\Delta_1\Delta_2}^{\Delta,2}$ admits a similar expression but with more complicated coefficients. We refrain from writing down the explicit formula but include it in the ancillary file accompanying the arXiv submission.  

The Polyakov-Regge blocks are linear combinations of the exchange and contact Witten diagrams \cite{Gimenez-Grau:2023fcy}
\begin{align}
 P^{\Delta,0}_{\Delta_1\Delta_2}
&= r_1 E^{\Delta,0}_{\Delta_1\Delta_2} \; , \\
 P^{\Delta,2}_{\Delta_1\Delta_2}
&= r_1 r_2 \left( E^{\Delta,2}_{\Delta_1\Delta_2}
 + r_3 C_{\Delta_1\Delta_2} \right) \, , \\
 \wh P^{\Dh,0}_{\Delta_1\Delta_2}
&= r_4 E^{\Dh,0}_{\Delta_1\Delta_2} \; , \quad
 \wh P^{\Dh,1}_{\Delta_1\Delta_2}
 = \chi\wh P^{\Dh,0}_{\Delta_1+1,\Delta_2+1} \; .
 \label{eq:polyregge}
\end{align}
The coefficients read
\begin{align*}
 r_1
&= \frac{2^{\Delta+2} (-1)^{\bar\Delta+1}
          \Gamma \left(\frac{\Delta +1}{2}\right)
          (1-\Delta_1)_{\bar\Delta}
          (1-\Delta_2)_{\bar\Delta}}
         {\pi^{3/2}
          \Gamma \! \left(\frac{\Delta -2}{2}\right)
          \left(\bar\Delta-1\right)!
          \left(\frac{8-\Delta -\Delta_1-\Delta_2}{2}\right)
             _{\bar\Delta-1}}\; , \\
 r_2
&= \frac{8 \left(\Delta ^2-1\right)}
        {(\Delta_1+\Delta_2-\Delta)
         (\Delta_1+\Delta_2+\Delta-6) (\Delta^2 -\Delta^2_{12})}
         \; , \\[0.4em]
 r_3
&=  \frac{7}{12}
  - \frac{\Delta_{12}^2 (\Delta_1+\Delta_2-6)^2}{12\Delta  (\Delta -6)} \\
& \quad
  + \frac{\left(\Delta_{12}^2-1\right) \left((\Delta_1+\Delta_2-6)^2-1\right)}{12(\Delta -1) (\Delta-5)} \; , \\
 r_4
&= \frac{8 \Gamma (\Delta_1) \Gamma (\Delta_2)}
        {\pi
         \Gamma (\frac{\Delta_1-\Dh}{2})
         \Gamma (\frac{\Delta_2-\Dh}{2} )
         \Gamma (\frac{\Delta_1+\Dh-2}{2} )
         \Gamma (\frac{\Delta_2+\Dh-2}{2} )} \; .
\end{align*}
Here we have chosen the normalization such that the conformal block of the exchanged field has unit coefficient.

In Mellin space, the Witten diagrams have simple expressions. The Mellin amplitude of a contact Witten diagram is just a constant
\begin{equation}
 \Mm_{C_{\Delta_1\Delta_2}}=\frac{\pi \Gamma\!\left(\frac{\Delta_1+\Delta_2-2}{2}\right)}{4 \Gamma(\Delta_1) \Gamma(\Delta_2)}\; .   
\end{equation}
For spin-0 and spin-2 exchange Witten diagrams in the bulk channel, we have
\begin{align}
\Mm_{\Delta_1\Delta_2}^{\Delta,0}
&= \sum_{n=0}^\infty \frac{R_n}{\delta - \frac{\Delta_1+\Delta_2-\Delta-2 n}{2}} \; , \\
 \Mm_{\Delta_1\Delta_2}^{\Delta,2}
&= \sum_{n=0}^\infty 
   \frac{S_n \gamma + T_n}{\delta - \frac{\Delta_1+\Delta_2-\Delta+2-2 n}{2}}
   + U \; .
 \label{eq:sumpolesblk}
\end{align}
The coefficients can be determined by imposing the equation of motion relation in Mellin space \cite{Chen:2023oax,Gimenez-Grau:2023fcy} and are given by 
\begin{align*}
 R_n
&= \frac{\pi
     \Gamma\!\left(\frac{\Delta-2}{2}\right)
     \Gamma\!\left(\frac{\Delta+\Delta_1+\Delta_2-6}{2}\right)
     \left(\frac{\Delta-2}{2}\right)_n \!
     \left(\frac{\Delta -\Delta_1-\Delta_2+2}{2}\right)_n}
    {16 n! \Gamma (\Delta_1) \Gamma (\Delta_2)
     \Gamma\!\left(\Delta+n-2\right)} \; , \\
 S_n 
&= \frac{2(\Delta -\Delta_1-\Delta_2) (\Delta +\Delta_1+\Delta_2-6)}{\Delta -\Delta_1-\Delta_2+2 n} R_n \; , \\
 T_n
&= \frac{S_n}{4 (\Delta -5) (\Delta -1) (\Delta +2 n-4)} \Big( \\
& \qquad  (\Delta -5) (\Delta -4) (\Delta -2) \Delta \\
& \qquad  + \Delta_{12}^2 \left(\Delta ^2-9 \Delta +2 n^2+2 \Delta  n-12 n+20\right) \\
& \qquad  + 2 \left(\Delta ^2-6 \Delta +4\right) n^2 \\
& \qquad  + 2 \left(\Delta ^3-8 \Delta ^2+16 \Delta -4\right) n \\
& \qquad  - 2 (\Delta -5) (\Delta -1) (\Delta_1+\Delta_2) (\Delta +2 n-4) \Big) \; , \\
 U
&= \frac{\pi \Gamma\!\left(\frac{\Delta_1+\Delta_2-2}{2}\right)}
        {48 (\Delta -5) (\Delta -1) \Gamma (\Delta_1) \Gamma (\Delta_2)} \bigg( \Delta_{12}^2 \\
& \qquad -7 \Delta (\Delta -6) +(\Delta_1 + \Delta_2-12) (\Delta_1 + \Delta_2) \\
& \qquad + 5\Delta_{12}^2 \, \frac{(\Delta_1 + \Delta_2)^2-12 (\Delta_1 + \Delta_2)+36}{(\Delta -6) \Delta } \, \bigg) \; .
\end{align*}
Note that the pole coefficients vanish for $n\geq \frac{\Delta_1+\Delta_2-\Delta+\ell}{2}$ when the quantum numbers satisfy $\Delta_1+\Delta_2-\Delta+\ell\in 2\mathbb{Z}_{>0}$. The exchange Mellin amplitudes then become rational functions. The situation for the defect exchange is similar
\begin{align*}
 \wh \Mm_{\Delta_1\Delta_2}^{\Dh,0}(\delta,\gamma)
&= \sum_{n=0}^\infty \frac{V_n}{\gamma-\Dh-2n} \; , \\
 \wh \Mm_{\Delta_1\Delta_2}^{\Dh,1}(\delta,\gamma)
&= 2 \Delta_1 \Delta_2 (\gamma-\delta)  \Mm_{\Delta_1+1,\Delta_2+1}^{\Dh,0}(\delta,\gamma +1) \; .
\end{align*}
The coefficients written in terms of $\Delta_{\text{d}i} = \Dh - \Delta_i$ are
\begin{align*}
 V_n
 = - \frac{\pi
           \Gamma\!\left(\frac{\Dh+\Delta_1-2}{2}\right)
           \Gamma\!\left(\frac{\Dh+\Delta_2-2}{2}\right)
           \left(\frac{2+\Delta_{\text{d}1}}{2}\right)_n
           \left(\frac{2+\Delta_{\text{d}2}}{2}\right)_n}
          {8 n! \Gamma (\Delta_1) \Gamma (\Delta_2) \Gamma(\Dh+n)} \; .
\end{align*}
The Mellin amplitude truncates to a rational function when $\Delta_1-\widehat{\Delta}+s\in 2\mathbb{Z}_{>0}$ or $\Delta_2-\widehat{\Delta}+s\in 2\mathbb{Z}_{>0}$.

\bibliography{refs} 

\providecommand{\href}[2]{#2}\begingroup\raggedright\begin{thebibliography}{10}

\bibitem{Wilson:1974sk}
K.~G. Wilson, ``{Confinement of Quarks},''
  \href{http://dx.doi.org/10.1103/PhysRevD.10.2445}{{\em Phys. Rev. D}
  {\bfseries 10} (1974) 2445--2459}.

\bibitem{Gaiotto:2014kfa}
D.~Gaiotto, A.~Kapustin, N.~Seiberg, and B.~Willett, ``{Generalized Global
  Symmetries},'' \href{http://dx.doi.org/10.1007/JHEP02(2015)172}{{\em JHEP}
  {\bfseries 02} (2015) 172}, \href{http://arxiv.org/abs/1412.5148}{{\ttfamily
  arXiv:1412.5148 [hep-th]}}.

\bibitem{Liendo:2012hy}
P.~Liendo, L.~Rastelli, and B.~C. van Rees, ``{The Bootstrap Program for
  Boundary CFT$_d$},'' \href{http://dx.doi.org/10.1007/JHEP07(2013)113}{{\em
  JHEP} {\bfseries 07} (2013) 113},
  \href{http://arxiv.org/abs/1210.4258}{{\ttfamily arXiv:1210.4258 [hep-th]}}.

\bibitem{Billo:2016cpy}
M.~Bill\`o, V.~Gon\c{c}alves, E.~Lauria, and M.~Meineri, ``{Defects in
  conformal field theory},''
  \href{http://dx.doi.org/10.1007/JHEP04(2016)091}{{\em JHEP} {\bfseries 04}
  (2016) 091}, \href{http://arxiv.org/abs/1601.02883}{{\ttfamily
  arXiv:1601.02883 [hep-th]}}.

\bibitem{Lemos:2017vnx}
M.~Lemos, P.~Liendo, M.~Meineri, and S.~Sarkar, ``{Universality at large
  transverse spin in defect CFT},''
  \href{http://dx.doi.org/10.1007/JHEP09(2018)091}{{\em JHEP} {\bfseries 09}
  (2018) 091}, \href{http://arxiv.org/abs/1712.08185}{{\ttfamily
  arXiv:1712.08185 [hep-th]}}.

\bibitem{Liendo:2019jpu}
P.~Liendo, Y.~Linke, and V.~Schomerus, ``{A Lorentzian inversion formula for
  defect CFT},'' \href{http://dx.doi.org/10.1007/JHEP08(2020)163}{{\em JHEP}
  {\bfseries 08} (2020) 163}, \href{http://arxiv.org/abs/1903.05222}{{\ttfamily
  arXiv:1903.05222 [hep-th]}}.

\bibitem{Bissi:2018mcq}
A.~Bissi, T.~Hansen, and A.~S\"oderberg, ``{Analytic Bootstrap for Boundary
  CFT},'' \href{http://dx.doi.org/10.1007/JHEP01(2019)010}{{\em JHEP}
  {\bfseries 01} (2019) 010}, \href{http://arxiv.org/abs/1808.08155}{{\ttfamily
  arXiv:1808.08155 [hep-th]}}.

\bibitem{Kaviraj:2018tfd}
A.~Kaviraj and M.~F. Paulos, ``{The Functional Bootstrap for Boundary CFT},''
  \href{http://dx.doi.org/10.1007/JHEP04(2020)135}{{\em JHEP} {\bfseries 04}
  (2020) 135}, \href{http://arxiv.org/abs/1812.04034}{{\ttfamily
  arXiv:1812.04034 [hep-th]}}.

\bibitem{Mazac:2018biw}
D.~Mazac, L.~Rastelli, and X.~Zhou, ``{An analytic approach to BCFT$_{d}$},''
  \href{http://dx.doi.org/10.1007/JHEP12(2019)004}{{\em JHEP} {\bfseries 12}
  (2019) 004}, \href{http://arxiv.org/abs/1812.09314}{{\ttfamily
  arXiv:1812.09314 [hep-th]}}.

\bibitem{Barrat:2022psm}
J.~Barrat, A.~Gimenez-Grau, and P.~Liendo, ``{A dispersion relation for defect
  CFT},'' \href{http://dx.doi.org/10.1007/JHEP02(2023)255}{{\em JHEP}
  {\bfseries 02} (2023) 255}, \href{http://arxiv.org/abs/2205.09765}{{\ttfamily
  arXiv:2205.09765 [hep-th]}}.

\bibitem{Bianchi:2022ppi}
L.~Bianchi and D.~Bonomi, ``{Conformal dispersion relations for defects and
  boundaries},'' \href{http://dx.doi.org/10.21468/SciPostPhys.15.2.055}{{\em
  SciPost Phys.} {\bfseries 15} no.~2, (2023) 055},
  \href{http://arxiv.org/abs/2205.09775}{{\ttfamily arXiv:2205.09775
  [hep-th]}}.

\bibitem{Die86a}
H.~W. Diehl, ``Field-theoretical approach to critical behaviour at surfaces,''
  in {\em Phase Transitions and Critical Phenomena}, C.~Domb and J.~L.
  Lebowitz, eds., vol.~10, pp.~75--267.
\newblock Academic, London, 1986.

\bibitem{cardy_1996}
J.~Cardy, \href{http://dx.doi.org/10.1017/CBO9781316036440}{{\em Scaling and
  Renormalization in Statistical Physics}}.
\newblock Cambridge Lecture Notes in Physics. Cambridge University Press, 1996.

\bibitem{McAvity:1995zd}
D.~M. McAvity and H.~Osborn, ``{Conformal field theories near a boundary in
  general dimensions},''
  \href{http://dx.doi.org/10.1016/0550-3213(95)00476-9}{{\em Nucl. Phys.}
  {\bfseries B455} (1995) 522--576},
\href{http://arxiv.org/abs/cond-mat/9505127}{{\ttfamily arXiv:cond-mat/9505127
  [cond-mat]}}.

\bibitem{Gaiotto:2013nva}
D.~Gaiotto, D.~Mazac, and M.~F. Paulos, ``{Bootstrapping the 3d Ising twist
  defect},'' \href{http://dx.doi.org/10.1007/JHEP03(2014)100}{{\em JHEP}
  {\bfseries 03} (2014) 100}, \href{http://arxiv.org/abs/1310.5078}{{\ttfamily
  arXiv:1310.5078 [hep-th]}}.

\bibitem{Cuomo:2021kfm}
G.~Cuomo, Z.~Komargodski, and M.~Mezei, ``{Localized magnetic field in the O(N)
  model},'' \href{http://dx.doi.org/10.1007/JHEP02(2022)134}{{\em JHEP}
  {\bfseries 02} (2022) 134}, \href{http://arxiv.org/abs/2112.10634}{{\ttfamily
  arXiv:2112.10634 [hep-th]}}.

\bibitem{Cuomo:2022xgw}
G.~Cuomo, Z.~Komargodski, M.~Mezei, and A.~Raviv-Moshe, ``{Spin impurities,
  Wilson lines and semiclassics},''
  \href{http://dx.doi.org/10.1007/JHEP06(2022)112}{{\em JHEP} {\bfseries 06}
  (2022) 112}, \href{http://arxiv.org/abs/2202.00040}{{\ttfamily
  arXiv:2202.00040 [hep-th]}}.

\bibitem{Giombi:2021uae}
S.~Giombi, E.~Helfenberger, Z.~Ji, and H.~Khanchandani, ``{Monodromy defects
  from hyperbolic space},''
  \href{http://dx.doi.org/10.1007/JHEP02(2022)041}{{\em JHEP} {\bfseries 02}
  (2022) 041}, \href{http://arxiv.org/abs/2102.11815}{{\ttfamily
  arXiv:2102.11815 [hep-th]}}.

\bibitem{Giombi:2020rmc}
S.~Giombi and H.~Khanchandani, ``{CFT in AdS and boundary RG flows},''
  \href{http://dx.doi.org/10.1007/JHEP11(2020)118}{{\em JHEP} {\bfseries 11}
  (2020) 118}, \href{http://arxiv.org/abs/2007.04955}{{\ttfamily
  arXiv:2007.04955 [hep-th]}}.

\bibitem{Bianchi:2022sbz}
L.~Bianchi, D.~Bonomi, and E.~de~Sabbata, ``{Analytic bootstrap for the
  localized magnetic field},''
  \href{http://dx.doi.org/10.1007/JHEP04(2023)069}{{\em JHEP} {\bfseries 04}
  (2023) 069}, \href{http://arxiv.org/abs/2212.02524}{{\ttfamily
  arXiv:2212.02524 [hep-th]}}.

\bibitem{Gimenez-Grau:2022ebb}
A.~Gimenez-Grau, ``{Probing magnetic line defects with two-point functions},''
  \href{http://arxiv.org/abs/2212.02520}{{\ttfamily arXiv:2212.02520
  [hep-th]}}.

\bibitem{Raviv-Moshe:2023yvq}
A.~Raviv-Moshe and S.~Zhong, ``{Phases of surface defects in Scalar Field
  Theories},'' \href{http://dx.doi.org/10.1007/JHEP08(2023)143}{{\em JHEP}
  {\bfseries 08} (2023) 143}, \href{http://arxiv.org/abs/2305.11370}{{\ttfamily
  arXiv:2305.11370 [hep-th]}}.

\bibitem{Trepanier:2023tvb}
M.~Tr\'epanier, ``{Surface defects in the O(N) model},''
  \href{http://dx.doi.org/10.1007/JHEP09(2023)074}{{\em JHEP} {\bfseries 09}
  (2023) 074}, \href{http://arxiv.org/abs/2305.10486}{{\ttfamily
  arXiv:2305.10486 [hep-th]}}.

\bibitem{Giombi:2023dqs}
S.~Giombi and B.~Liu, ``{Notes on a Surface Defect in the $O(N)$ Model},''
  \href{http://arxiv.org/abs/2305.11402}{{\ttfamily arXiv:2305.11402
  [hep-th]}}.

\bibitem{Chiodaroli:2016jod}
M.~Chiodaroli, J.~Estes, and Y.~Korovin, ``{Holographic two-point functions for
  Janus interfaces in the $D1/D5$ CFT},''
  \href{http://dx.doi.org/10.1007/JHEP04(2017)145}{{\em JHEP} {\bfseries 04}
  (2017) 145}, \href{http://arxiv.org/abs/1612.08916}{{\ttfamily
  arXiv:1612.08916 [hep-th]}}.

\bibitem{Giombi:2017cqn}
S.~Giombi, R.~Roiban, and A.~A. Tseytlin, ``{Half-BPS Wilson loop and
  AdS$_2$/CFT$_1$},''
  \href{http://dx.doi.org/10.1016/j.nuclphysb.2017.07.004}{{\em Nucl. Phys. B}
  {\bfseries 922} (2017) 499--527},
  \href{http://arxiv.org/abs/1706.00756}{{\ttfamily arXiv:1706.00756
  [hep-th]}}.

\bibitem{Drukker:2020swu}
N.~Drukker, S.~Giombi, A.~A. Tseytlin, and X.~Zhou, ``{Defect CFT in the 6d
  (2,0) theory from M2 brane dynamics in AdS$_7 \times$S$^4$},''
  \href{http://dx.doi.org/10.1007/JHEP07(2020)101}{{\em JHEP} {\bfseries 07}
  (2020) 101}, \href{http://arxiv.org/abs/2004.04562}{{\ttfamily
  arXiv:2004.04562 [hep-th]}}.

\bibitem{Ferrero:2021bsb}
P.~Ferrero and C.~Meneghelli, ``{Bootstrapping the half-BPS line defect CFT in
  N=4 supersymmetric Yang-Mills theory at strong coupling},''
  \href{http://dx.doi.org/10.1103/PhysRevD.104.L081703}{{\em Phys. Rev. D}
  {\bfseries 104} no.~8, (2021) L081703},
  \href{http://arxiv.org/abs/2103.10440}{{\ttfamily arXiv:2103.10440
  [hep-th]}}.

\bibitem{Barrat:2021yvp}
J.~Barrat, A.~Gimenez-Grau, and P.~Liendo, ``{Bootstrapping holographic defect
  correlators in $ \mathcal{N} $ = 4 super Yang-Mills},''
  \href{http://dx.doi.org/10.1007/JHEP04(2022)093}{{\em JHEP} {\bfseries 04}
  (2022) 093}, \href{http://arxiv.org/abs/2108.13432}{{\ttfamily
  arXiv:2108.13432 [hep-th]}}.

\bibitem{Meneghelli:2022gps}
C.~Meneghelli and M.~Tr\'epanier, ``{Bootstrapping string dynamics in the 6d
  $\mathcal{N}$ = (2, 0) theories},''
  \href{http://dx.doi.org/10.1007/JHEP07(2023)165}{{\em JHEP} {\bfseries 07}
  (2023) 165}, \href{http://arxiv.org/abs/2212.05020}{{\ttfamily
  arXiv:2212.05020 [hep-th]}}.

\bibitem{Gimenez-Grau:2023fcy}
A.~Gimenez-Grau, ``{The Witten Diagram Bootstrap for Holographic Defects},''
  \href{http://arxiv.org/abs/2306.11896}{{\ttfamily arXiv:2306.11896
  [hep-th]}}.

\bibitem{Giombi:2023zte}
S.~Giombi, S.~Komatsu, B.~Offertaler, and J.~Shan, ``{Boundary
  reparametrizations and six-point functions on the AdS$_2$ string},''
  \href{http://arxiv.org/abs/2308.10775}{{\ttfamily arXiv:2308.10775
  [hep-th]}}.

\bibitem{Rastelli:2016nze}
L.~Rastelli and X.~Zhou, ``{Mellin amplitudes for $AdS_5\times S^5$},''
  \href{http://dx.doi.org/10.1103/PhysRevLett.118.091602}{{\em Phys. Rev.
  Lett.} {\bfseries 118} no.~9, (2017) 091602},
\href{http://arxiv.org/abs/1608.06624}{{\ttfamily arXiv:1608.06624 [hep-th]}}.

\bibitem{Rastelli:2017udc}
L.~Rastelli and X.~Zhou, ``{How to Succeed at Holographic Correlators Without
  Really Trying},'' \href{http://dx.doi.org/10.1007/JHEP04(2018)014}{{\em JHEP}
  {\bfseries 04} (2018) 014},
\href{http://arxiv.org/abs/1710.05923}{{\ttfamily arXiv:1710.05923 [hep-th]}}.

\bibitem{Alday:2023kfm}
L.~F. Alday, V.~Gon\c{c}alves, M.~Nocchi, and X.~Zhou, ``{Six-Point AdS Gluon
  Amplitudes from Flat Space and Factorization},''
  \href{http://arxiv.org/abs/2307.06884}{{\ttfamily arXiv:2307.06884
  [hep-th]}}.

\bibitem{Huang:2021xws}
Z.~Huang and E.~Y. Yuan, ``{Graviton scattering in AdS$_{5}$\texttimes{}
  S$^{5}$ at two loops},''
  \href{http://dx.doi.org/10.1007/JHEP04(2023)064}{{\em JHEP} {\bfseries 04}
  (2023) 064}, \href{http://arxiv.org/abs/2112.15174}{{\ttfamily
  arXiv:2112.15174 [hep-th]}}.

\bibitem{Huang:2023oxf}
Z.~Huang, B.~Wang, E.~Y. Yuan, and X.~Zhou, ``{AdS super gluon scattering up to
  two loops: a position space approach},''
  \href{http://dx.doi.org/10.1007/JHEP07(2023)053}{{\em JHEP} {\bfseries 07}
  (2023) 053}, \href{http://arxiv.org/abs/2301.13240}{{\ttfamily
  arXiv:2301.13240 [hep-th]}}.

\bibitem{Bissi:2022mrs}
A.~Bissi, A.~Sinha, and X.~Zhou, ``{Selected topics in analytic conformal
  bootstrap: A guided journey},''
  \href{http://dx.doi.org/10.1016/j.physrep.2022.09.004}{{\em Phys. Rept.}
  {\bfseries 991} (2022) 1--89},
  \href{http://arxiv.org/abs/2202.08475}{{\ttfamily arXiv:2202.08475
  [hep-th]}}.

\bibitem{Note1}
One can also consider $M$ coincident M2-branes but this will only change the
  results of the paper by an overall factor.

\bibitem{Note2}
The normalization is ${\delimiter "426830A S_k S_k\delimiter "526930B
  }_{\protect \rm free}=(u_1\cdot u_2)^k/x_{12}^{4k}$ in (\ref {SSVexpand}).

\bibitem{bllprv13}
C.~Beem, M.~Lemos, P.~Liendo, W.~Peelaers, L.~Rastelli, and B.~C. van Rees,
  ``{Infinite chiral symmetry in four dimensions},''
  \href{http://dx.doi.org/10.1007/s00220-014-2272-x}{{\em Commun. Math. Phys.}
  {\bfseries 336} (2015) 1359--1433},
  \href{http://arxiv.org/abs/1312.5344}{{\ttfamily 1312.5344}}.

\bibitem{brv14}
C.~Beem, L.~Rastelli, and B.~C. van Rees, ``{W symmetry in six dimensions},''
  \href{http://dx.doi.org/10.1007/JHEP05(2017)017}{{\em JHEP} {\bfseries 05}
  (2017) 017}, \href{http://arxiv.org/abs/1404.1079}{{\ttfamily 1404.1079}}.

\bibitem{Note3}
More precisely, the bulk field transforms in the rank-$R_{\protect \mathcal
  {X}}$ symmetric traceless representation of the original $SO(5)$ R-symmetry
  group while the defect field transforms in the rank-$R_{\protect \mathcal
  {Y}}$ representation with respect to the unbroken $SO(4)$ R-symmetry.

\bibitem{Rastelli:2017ecj}
L.~Rastelli and X.~Zhou, ``{The Mellin Formalism for Boundary CFT$_d$},''
  \href{http://dx.doi.org/10.1007/JHEP10(2017)146}{{\em JHEP} {\bfseries 10}
  (2017) 146},
\href{http://arxiv.org/abs/1705.05362}{{\ttfamily arXiv:1705.05362 [hep-th]}}.

\bibitem{Corrado:1999pi}
R.~Corrado, B.~Florea, and R.~McNees, ``{Correlation functions of operators and
  Wilson surfaces in the d = 6, (0,2) theory in the large N limit},''
  \href{http://dx.doi.org/10.1103/PhysRevD.60.085011}{{\em Phys. Rev. D}
  {\bfseries 60} (1999) 085011},
  \href{http://arxiv.org/abs/hep-th/9902153}{{\ttfamily arXiv:hep-th/9902153}}.

\bibitem{Bastianelli:1999en}
F.~Bastianelli and R.~Zucchini, ``{Three point functions of chiral primary
  operators in d = 3, N=8 and d = 6, N=(2,0) SCFT at large N},''
  \href{http://dx.doi.org/10.1016/S0370-2693(99)01179-X}{{\em Phys. Lett. B}
  {\bfseries 467} (1999) 61--66},
  \href{http://arxiv.org/abs/hep-th/9907047}{{\ttfamily arXiv:hep-th/9907047}}.

\bibitem{Note4}
Our result for $a_k$ agrees with \cite {Corrado:1999pi} up to a simple factor.
  A similar mismatch was also observed in \cite {Bastianelli:1999en} for
  three-point functions.

\bibitem{Rastelli:2017ymc}
L.~Rastelli and X.~Zhou, ``{Holographic Four-Point Functions in the (2, 0)
  Theory},'' \href{http://dx.doi.org/10.1007/JHEP06(2018)087}{{\em JHEP}
  {\bfseries 06} (2018) 087}, \href{http://arxiv.org/abs/1712.02788}{{\ttfamily
  arXiv:1712.02788 [hep-th]}}.

\bibitem{Behan:2021pzk}
C.~Behan, P.~Ferrero, and X.~Zhou, ``{More on holographic correlators: Twisted
  and dimensionally reduced structures},''
  \href{http://dx.doi.org/10.1007/JHEP04(2021)008}{{\em JHEP} {\bfseries 04}
  (2021) 008}, \href{http://arxiv.org/abs/2101.04114}{{\ttfamily
  arXiv:2101.04114 [hep-th]}}.

\bibitem{Mack:2009mi}
G.~Mack, ``{D-independent representation of Conformal Field Theories in D
  dimensions via transformation to auxiliary Dual Resonance Models. Scalar
  amplitudes},''
\href{http://arxiv.org/abs/0907.2407}{{\ttfamily arXiv:0907.2407 [hep-th]}}.

\bibitem{Penedones:2010ue}
J.~Penedones, ``{Writing CFT correlation functions as AdS scattering
  amplitudes},'' \href{http://dx.doi.org/10.1007/JHEP03(2011)025}{{\em JHEP}
  {\bfseries 03} (2011) 025},
\href{http://arxiv.org/abs/1011.1485}{{\ttfamily arXiv:1011.1485 [hep-th]}}.

\bibitem{Goncalves:2018fwx}
V.~Goncalves and G.~Itsios, ``{A note on defect Mellin amplitudes},''
  \href{http://arxiv.org/abs/1803.06721}{{\ttfamily arXiv:1803.06721
  [hep-th]}}.

\bibitem{Caron-Huot:2018kta}
S.~Caron-Huot and A.-K. Trinh, ``{All Tree-Level Correlators in
  AdS${}_5\times$S${}_5$ Supergravity: Hidden Ten-Dimensional Conformal
  Symmetry},'' \href{http://dx.doi.org/10.1007/JHEP01(2019)196}{{\em JHEP}
  {\bfseries 01} (2019) 196},
\href{http://arxiv.org/abs/1809.09173}{{\ttfamily arXiv:1809.09173 [hep-th]}}.

\bibitem{Rastelli:2019gtj}
L.~Rastelli, K.~Roumpedakis, and X.~Zhou, ``{$\mathbf{AdS_3\times S^3}$
  Tree-Level Correlators: Hidden Six-Dimensional Conformal Symmetry},''
  \href{http://dx.doi.org/10.1007/JHEP10(2019)140}{{\em JHEP} {\bfseries 10}
  (2019) 140},
\href{http://arxiv.org/abs/1905.11983}{{\ttfamily arXiv:1905.11983 [hep-th]}}.

\bibitem{Alday:2021odx}
L.~F. Alday, C.~Behan, P.~Ferrero, and X.~Zhou, ``{Gluon Scattering in AdS from
  CFT},'' \href{http://dx.doi.org/10.1007/JHEP06(2021)020}{{\em JHEP}
  {\bfseries 06} (2021) 020}, \href{http://arxiv.org/abs/2103.15830}{{\ttfamily
  arXiv:2103.15830 [hep-th]}}.

\bibitem{Abl:2021mxo}
T.~Abl, P.~Heslop, and A.~E. Lipstein, ``{Higher-dimensional symmetry of
  AdS$_{2}$\texttimes{}S$^{2}$ correlators},''
  \href{http://dx.doi.org/10.1007/JHEP03(2022)076}{{\em JHEP} {\bfseries 03}
  (2022) 076}, \href{http://arxiv.org/abs/2112.09597}{{\ttfamily
  arXiv:2112.09597 [hep-th]}}.

\bibitem{Alday:2020lbp}
L.~F. Alday and X.~Zhou, ``{All Tree-Level Correlators for M-theory on $AdS_7
  \times S^4$},'' \href{http://dx.doi.org/10.1103/PhysRevLett.125.131604}{{\em
  Phys. Rev. Lett.} {\bfseries 125} no.~13, (2020) 131604},
  \href{http://arxiv.org/abs/2006.06653}{{\ttfamily arXiv:2006.06653
  [hep-th]}}.

\bibitem{Aharony:2016dwx}
O.~Aharony, L.~F. Alday, A.~Bissi, and E.~Perlmutter, ``{Loops in AdS from
  Conformal Field Theory},''
  \href{http://dx.doi.org/10.1007/JHEP07(2017)036}{{\em JHEP} {\bfseries 07}
  (2017) 036},
\href{http://arxiv.org/abs/1612.03891}{{\ttfamily arXiv:1612.03891 [hep-th]}}.

\bibitem{Alday:2020dtb}
L.~F. Alday and X.~Zhou, ``{All Holographic Four-Point Functions in All
  Maximally Supersymmetric CFTs},''
  \href{http://dx.doi.org/10.1103/PhysRevX.11.011056}{{\em Phys. Rev. X}
  {\bfseries 11} no.~1, (2021) 011056},
  \href{http://arxiv.org/abs/2006.12505}{{\ttfamily arXiv:2006.12505
  [hep-th]}}.

\bibitem{Chen:2023oax}
J.~Chen and X.~Zhou, ``{Aspects of higher-point functions in BCFT$_{d}$},''
  \href{http://dx.doi.org/10.1007/JHEP09(2023)204}{{\em JHEP} {\bfseries 09}
  (2023) 204}, \href{http://arxiv.org/abs/2304.11799}{{\ttfamily
  arXiv:2304.11799 [hep-th]}}.

\end{thebibliography}\endgroup
\bibliographystyle{utphys}
\end{document}